\documentclass{article}
\usepackage{spconf,amsmath,graphicx}
\usepackage{stfloats}
\usepackage{changepage}
\usepackage{color}
\usepackage[normalem]{ ulem }
\usepackage{soul}
\title{RADIOGAN: Deep Convolutional Conditional Generative adversarial Network To Generate PET Images}
\name{A. Amyar \textsuperscript{1, 2}, S. Ruan \textsuperscript{2} , 
 P. Vera \textsuperscript{2, 3}, P. Decazes \textsuperscript{2, 3},
R. Modzelewski \textsuperscript{2, 3}}
\address{\textsuperscript{1} General Electric Healthcare, Buc, France \\
\textsuperscript{2} LITIS -  EA4108 - Quantif, University of Rouen, Rouen, France  \\
\textsuperscript{3} Nuclear Medicine Department, Henri Becquerel Center, Rouen, France
}

\begin{document}

\maketitle

\begin{abstract}

One of the most challenges in medical imaging is the lack of data. It is proven that classical data augmentation methods are useful but still limited due to the huge variation in images. Using generative adversarial networks (GAN) is a promising way to address this problem, however, it is challenging to train one model to generate different classes of lesions. In this paper, we propose a deep convolutional conditional generative adversarial network to generate MIP positron emission tomography image (PET) which is a 2D image that represents a 3D volume for fast interpretation, according to different lesions or non lesion (normal). The advantage of our proposed method consists of one model that is capable of generating different classes of lesions trained on a small sample size for each class of lesion, and showing a very promising results.
In addition, we show that a walk through a latent space can be used as a tool to evaluate the images generated.

\end{abstract}

\begin{keywords}
Machine Learning, Deep learning, Generative adversarial Networks, Positron Emission Tomography
\end{keywords}

\section{Introduction}

A generative adversarial network (GAN)  is a class of machine learning system invented by Ian Goodfellow in 2014 \cite{goodfellow2014generative}. Two neural networks a discriminator and a generator compete with each other in a  mini-max game. Given a training set, the generator tends to learn how to produce new data with the same statistics as the training set, while the discriminator tries to distinguish between real data and generated one as shown in Fig. 1. Generative methods (in particular, GANs) are currently used in various places for data augmentation \cite{perez2017effectiveness}. Their potential is vast; they can learn to mimic any distribution of data across any domain: photographs, drawings, music, and prose. However, there’s no ground truth data to predict, which make it difficult to measure the model performance, especially in the medical field where the images may contain different lesions. In addition, it is not evident to distinguish a memorizing GAN, which memorize training examples from a generating one that learned meaningful boundary between real and generated images \cite{brock2018large}. Walking through a latent space can be used as a tool that can evaluate memorization degree.

\begin{figure}

		\includegraphics[height=4cm, width=8cm ]{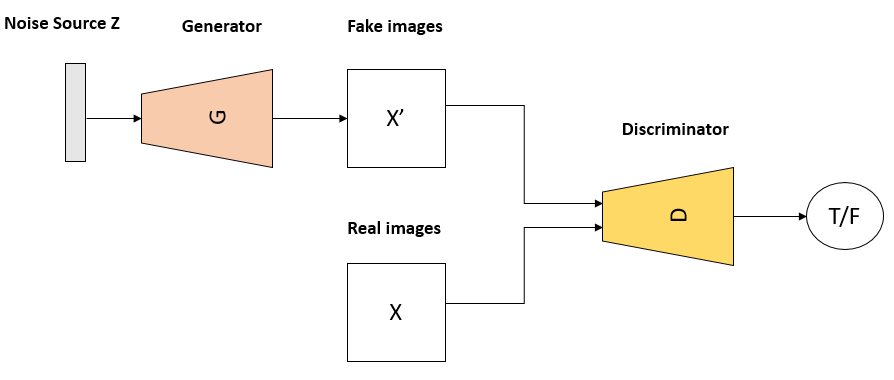}
		\caption{Generative Adversarial Network Framework. }

   \label{fig::cancer1}
\end{figure}

Several authors have used GANs to build models for medical imaging synthesis, such as generating computed tomography (CT) from magnetic resonance imaging (MRI) \cite{nie2017medical} or learning with GANs for chest X-ray classification \cite{madani2018semi}. In PET images, Ben-Cohen et al. \cite{ben2017virtual} proposed a framework to generate PET images from CT data using a deep convolutional network. However, to our knowledge, there was no study about the generation of different class of lesions with the same model, and neither the verification if the model able to generate a complete new data or its memorizing the images from the training set.

To perform fast lesion classification physicians usually use MIP PET images which are 2D images that represents the 3D images. In this paper, we focus on the generation of MIP PET images. Deep learning methods trained to perform such a task are data hungry and need a lot of data to accomplish this task due the variability in images with different lesions. GANs are a promising tool that can solve this problem by increasing the size of data. However, we are faced with two main challenges when training a GAN: first, each class of lesions presents a small dataset in its own when used separately. Second, it is usually hard to verify if the GAN is generalizing a complete new data or it is just memorizing the dataset. Moreover, medical imaging presents a unique class of difficulties due to the huge variability in images, the presence of lesions in a specific anatomical point, the lack of integration of medical information and other constraints. 

In this work, we develop a new GAN architecture, called RadioGAN, to generate FDG-PET images of normal patients and lesions. The idea is to combine different classes of lesions to train a condtional model which can capture not only common features in PET images but also specific ones for each lesion. We  address both issues of data generation and class specified data augmentation. Moreover, we propose a framework to evaluate the result of a GAN and to verify if it is generating a complete new images or if it is memorizing the images seen in training by using a walk through a latent space \cite{radford2015unsupervised}. 
If a GAN memorizes images, then choosing random seeds in latent space creates basic blends of training images. However, if a GAN generalizes good images, then choosing random seeds produces exciting images that utilize patterns and components from the training dataset but are not simple blend. 

\begin{figure*}

		\centering
		\includegraphics[height=6.5cm, width=14cm]{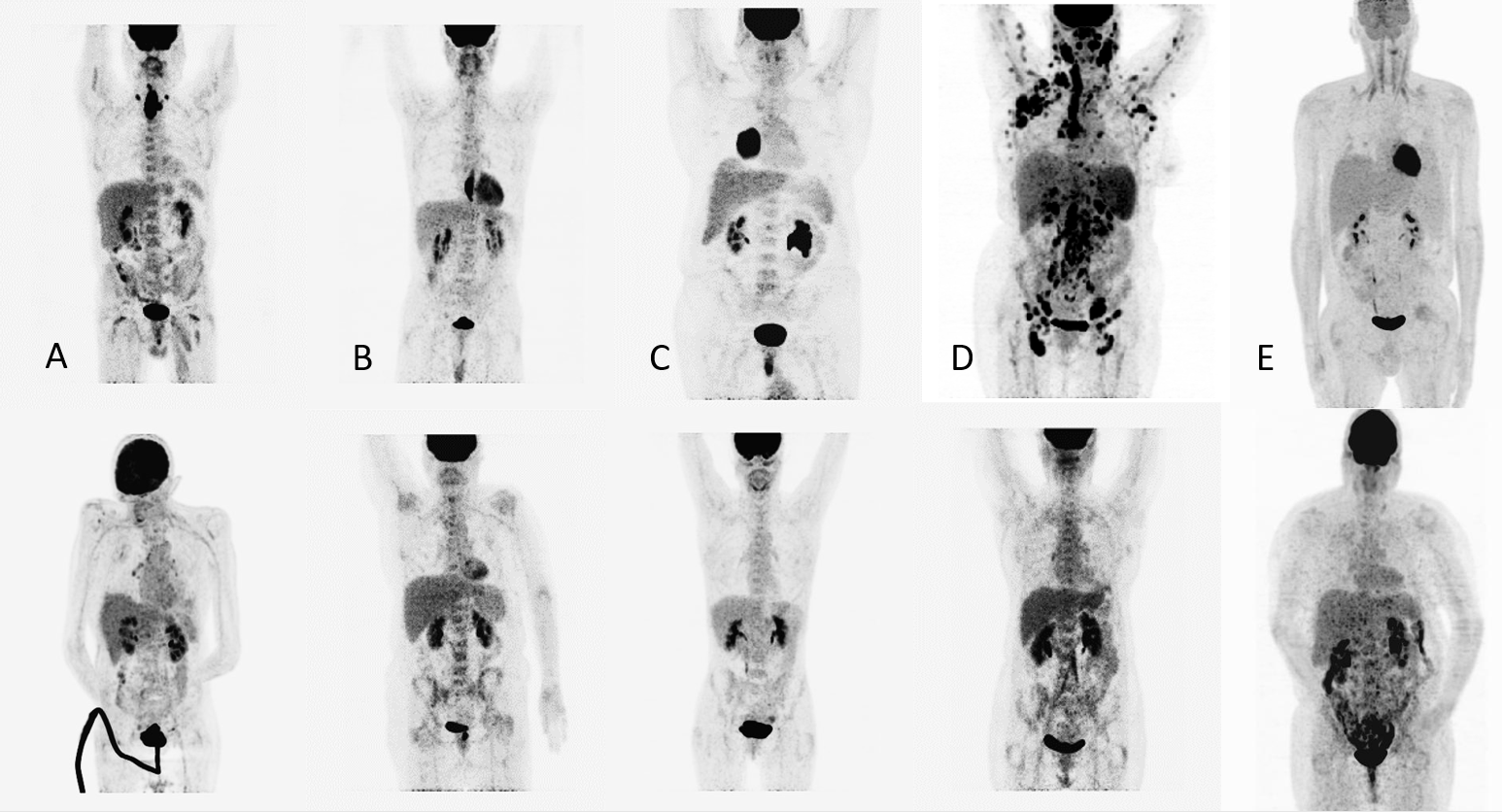}
		\caption{Examples of PET-MIP input images used to train GAN. In the upper line from left to right: A: head \& neck cancer, B: oesophagus cancer, C: lung cancer, D: lymphoma and E: normal patient. In the second line,  showing multiple cases for the same category (normal).}

   \label{fig::cancer2}
\end{figure*}

\section{Data}

We have one thousand six hundred and six patients (1606) in which six hundred and seventy five (675) are normal (only physiological uptakes) and nine hundred and thirty one (931) with pathological uptakes (cancer). We have four classes of cancer: lymphoma (225), lung cancer (189), oesophagus cancer (97) and head and neck cancer (422). Patient information was anonymized prior for analysis. The voxel size of  PET exams is $4.06\times4.06\times2.0$  mm\textsuperscript{3} with a size of 429 x 168 x 168 before preprocessing. The lung cancer and head \& neck datasets are from the cancer imaging archive (TCIA) \cite{clark2013cancer}. The images were  acquired at the Henri Becquerel Center Institution and the study was approved as a retrospective study.

\section{METHOD}
The object is to generate a specific class of image from one model trained with multiple classes (conditions).
\subsection{Data preprocessing}
Images were spatially normalized by resampling all the dataset to an isotropic resolution of $2\times2\times2$ mm\textsuperscript{3} using the k-nearest neighbor interpolation algorithm. PET images gray level intensity were normalized to absolute Standardized Uptake Value (SUV) level between [0 30] and translated between [0 1] to be used in GAN architectures. Maximum intensity Projection (MIP) transformation was used to compress a 3D exam into a 2D representation. Five classes: normal, head \& neck cancer, oesophagus cancer, lung cancer and lymphoma are shown in Figure 2. 

\subsection{Proposed RadioGAN architecture}
RadioGAN is inspired from deep convolutional generative adversarial network (DCGAN) which is a class of convolutional neural networks that generate images in an unsupervised learning way \cite{radford2015unsupervised} and conditional generative adversarial network (CGAN) which is a conditional version of GAN constructed by feeding the GAN the data and the label to condition both the generator and the discriminator \cite{mirza2014conditional}.

The design of a GAN can be a challenging part, especially the generator design. The Generator takes as input a random noise and then map it into an image such that the discriminator cannot tell which images came from the dataset (real) and which ones came from the generator (fake).Several studies showed that using Convolutional Layers in the generator network produces better results \cite{radford2015unsupervised}. To transform a GAN to a DCGAN we follow the steps as in \cite{radford2015unsupervised}: replace all max pooling with convolutional stride, use transposed convolution for upsampling, eliminate fully connected layers, use of batch normalization except for the output layer for the generator and the input layer of the discriminator, use ReLU in the generator except for the output which uses tanh and LeakyReLU in the discriminator.

Conditional GAN uses image class information as a condition. For example, when we give our Generator a random vector Z, at the same time, we also give the class label Y= [Normal, Esophagus cancer, head \& neck cancer, lung cancer, lymphoma]. Then the Discriminator will evaluate how well the Generator draws a patient of the corresponding class Label. After that, we use the evaluated error to update the Generator's weights (which includes new weights to accommodate the input Y).

\begin{figure}

	    \centering
		\includegraphics[height=6cm, width=8cm]{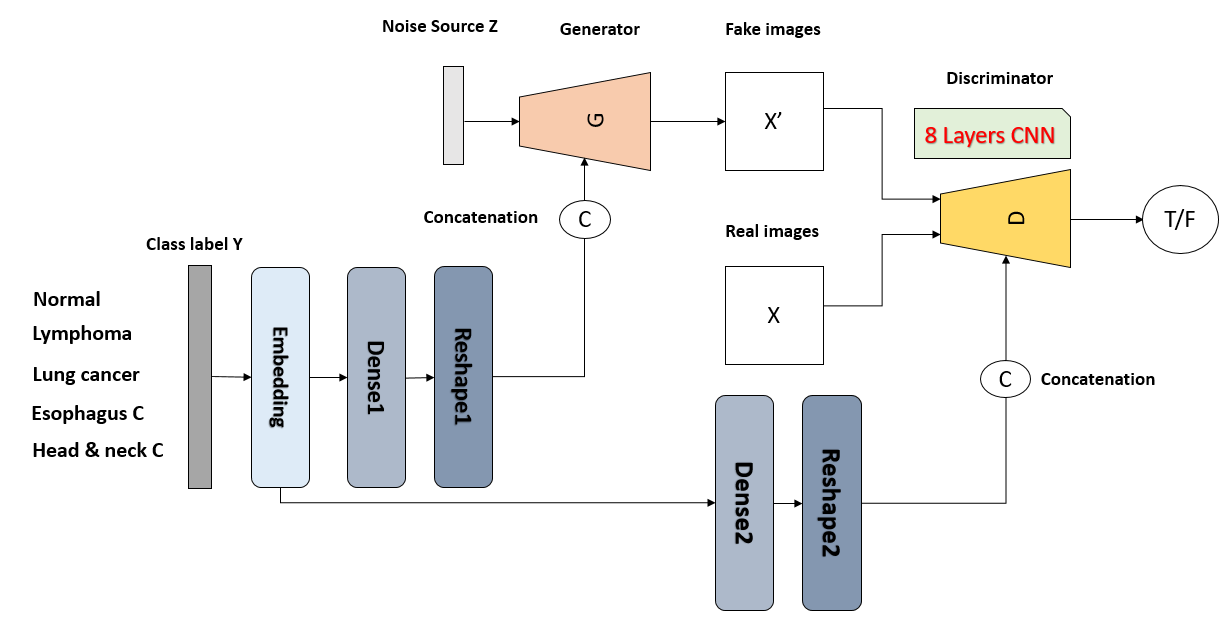}
		\caption{RadioGANarchitecture}

   \label{fig::cancer3}
\end{figure}

In RadioGAN we take advantage of  both DCGAN and CGAN to create a novel architecture DCCGAN (Deep Convolutional Conditional Generative adversarial Network) to generate MIP PET images for different classes of cancer and also for normal patient. The class label is added as an additional input layer for both generator and discriminator. To specify the class of the lesion for the generator, we encoded it as a condition using an embedding layer, which is a dense vector of a fixed size of 50. The five classes of images: lung cancer, esophagus cancer, head \& neck cancer, lymphoma and normal are mapped to a different 50 element vector representation. Then, the output is followed by a Dense layer with a linear activation. The Dense layer has 15 neurons to match the 5 x 3 feature map activation of the generator model, and added as a second channel. For the discriminator, a new Embedding layer of a size of 50 is added, followed by a Dense layer to scale up to image dimension with a linear activation and added to the input image as a second channel as shown in Fig. 3.
 
The discriminator is an eight layers deep convolutional neural network in which each layer is composed of a convolution with a filter of 3 X 3 and a stride of 2 followed by a LeakyReLU and a Dropout with no pooling, and a BatchNormalization with a momentum of 0.8 except for the first and the two last layers. The last layer is a Dense layer with a Sigmoid function to classify the image for real or fake. The generator takes a random noise Z and passed to a Dense layer, followed by a Reshape and a BatchNormalization of a momentum of 0.8. Then a series of transposed convolution followed by a Batchnormalization are applied except for the last layer which is a convolutional layer with one channel and a tanh activation function. Then, the adversarial network is extended to a conditional model by conditioning both the discriminator and the generator on the lesions, normal labels. 

\begin{figure}

		\includegraphics[height=7cm, width=8cm ]{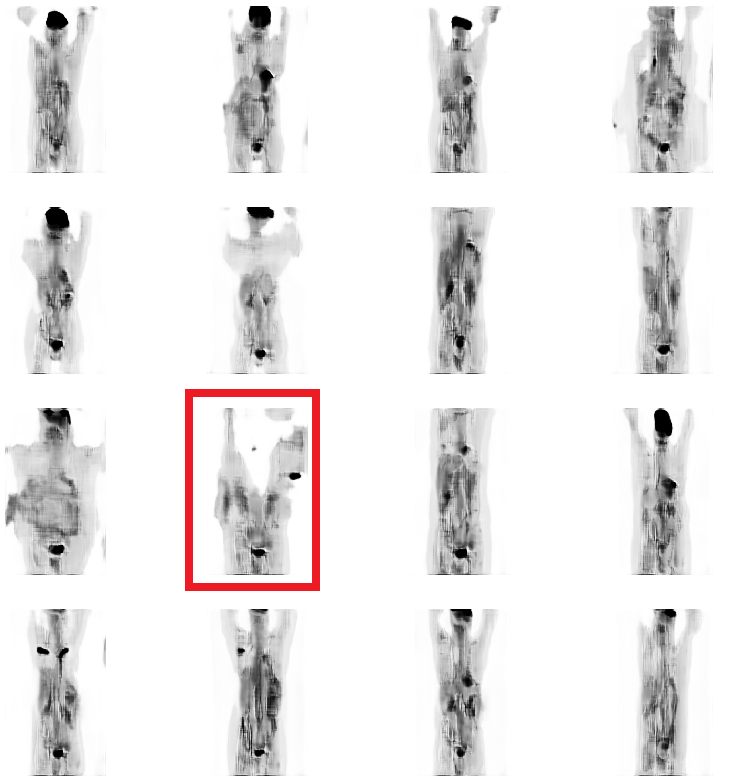}
		\caption{An example of a memorizing GAN. The first seed is for a Normal patient and the last one for head \& neck cancer. The GAN model is blending images between the two seeds to go from a normal patient to head \& neck cancer. The image framed in red is not realistic.}

   \label{fig::cancer4}
\end{figure}

We have trained our model for 300 epochs and used the binary crossentropy loss and Adam optimizer with a learning rate of 0.0002 and beta1 of 0.5. Accuracy is used as the metric for the discriminator.
 
 \section{Results}
\subsection{Evaluation of the memorization vs generalisation}
Once the model is obtained, we can input a random vector (seed) of length 100 into the Generator and we get a patient image. If we input seed Z1 = [1, 0, 0...0] for example, we may get a patient with oesophagus cancer image and if we input seed Z10 = [10, 0, 0...0] we may obtain a normal patient image. What happens when we input the seeds between Z1 and Z10 ? For example Z2 = [2, 0, 0...0], Z3 = [3, 0, 0...0], … , Z9 = [9, 0, 0...0] ? A high quality generator must always output realistic images, so we should not see a half patient, a patient with 3 legs or a head instead of the stomach. Instead, we should see a sequence of realistic patient images that slowly transform from oesophagus cancer to normal. Latent walks can help us to distinguish simple memorizing GANs from complex generalizing GANs. In Fig. 4 an example of images generated with a standard CGAN which tends to memorize, we can see the image framed in red is not realistic. While in Fig. 5 an example of images generated with our proposed RADIOGAN which tends to generate realistic new images and not only blending them.

\begin{figure*}

		\centering
		\includegraphics[height=7cm, width=17cm]{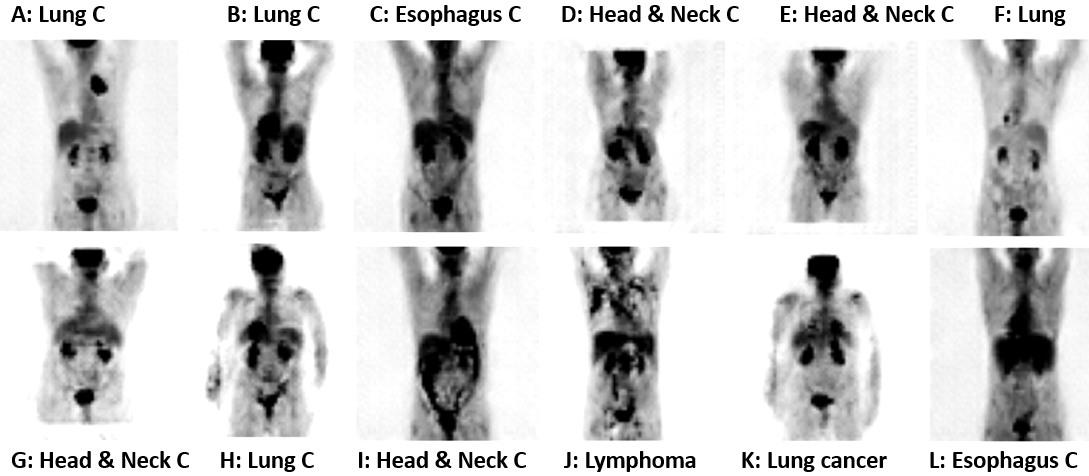}
		\caption{An example of images generated with RadioGAN. The model is moving slowly from lung cancer (A) to lymphoma (J) while generating a sequence of realistic patient images with coherence..}

   \label{fig::cancer5}
\end{figure*}

\section{Discussion and Conclusion}
We have developed an end-to-end convolutional conditional generative adversarial network to generate MIP PET images. The advantage of this strategy is to train one model which can capture common features, but also specific ones for each lesion and for normal class. The integration of class lesions and normal class information is an interesting and challenging approach due to the difficulty to evaluate a GAN model. We have approached this problem through a walk in the latent space to evaluate the quality of the images generated.  

Generative adversarial networks are a very promising tool to tackle the lack of data in the medical imaging field, and it can be also used to provide an advantage for other problems such as domain adaptation where data come from different distribution.  RadioGAN shows promising results and can be extended to whole 3D body images to generate PET images from PET for data augmentation or from another modality such as CT for data synthesis. These results can be confirmed on a larger database.

\section{Acknowledgments}

The authors acknowledge financial support from the Canceropole Nord-Ouest for emerging projects \\ (https://www.canceropole-nordouest.org/).

{\footnotesize
\bibliographystyle{IEEEbib}
\bibliography{main.bib}
}
\end{document}